\documentclass{article}
\usepackage[utf8]{inputenc}
\usepackage[normalem]{ulem}
\usepackage{graphicx} 
\usepackage{authblk}
\usepackage{xcolor}
\usepackage{tablefootnote}
\usepackage{cite}
\usepackage{amsmath,amssymb,amsfonts}
\usepackage{mathtools}
\usepackage{algorithmic}
\usepackage{textcomp}
\usepackage{relsize}

\usepackage[font=small,labelfont=bf]{caption}
\usepackage[square,numbers]{natbib}
\usepackage{hyperref}

\usepackage{graphicx}
\usepackage{dcolumn}
\usepackage{bm}

\usepackage[utf8]{inputenc}
\usepackage[T1]{fontenc}
\usepackage{mathptmx}
\usepackage{etoolbox}
\usepackage{xcolor}
\usepackage[table]{xcolor}
\usepackage{multirow}
\usepackage{booktabs}
\usepackage{mathrsfs}
\usepackage[mathcal]{euscript}
\usepackage{cite}
\usepackage{makecell}
\usepackage{siunitx}

\newcounter{bla}

\newcommand{\be}{\begin{equation}}

\newcommand{\ee}{\end{equation}}
\definecolor{codegreen}{rgb}{0,0.6,0}
\definecolor{codegray}{rgb}{0.5,0.5,0.5}
\definecolor{codepurple}{rgb}{0.58,0,0.82}
\definecolor{backcolour}{rgb}{0.95,0.95,0.92}

\usepackage{listings}
\definecolor{mGreen}{rgb}{0,0.6,0}
\definecolor{mGray}{rgb}{0.5,0.5,0.5}
\definecolor{mPurple}{rgb}{0.58,0,0.82}
\definecolor{backgroundColour}{rgb}{0.95,0.95,0.92}
\lstdefinestyle{CStyle}{
    backgroundcolor=\color{backgroundColour},   
    commentstyle=\color{mGreen},
    keywordstyle=\color{magenta},
    numberstyle=\tiny\color{mGray},
    stringstyle=\color{mPurple},
    basicstyle=\footnotesize,
    breakatwhitespace=false,         
    breaklines=true,                 
    captionpos=b,                    
    keepspaces=true,                 
    numbers=left,                    
    numbersep=5pt,                  
    showspaces=false,                
    showstringspaces=false,
    showtabs=false,                  
    tabsize=2,
    language=C
}

\title{Interpretable statistical feature engineering for early disruption prediction in the short pulse ADITYA tokamak}
\author{Jyoti Agarwal$^{1,2,*}$, Kavit Patel$^1$, Bhaskar Chaudhury$^{1,*}$, Abhishek Sharma$^2$, Shrichand Jakhar$^2$, Manika Sharma$^2$}

\affil{$^1$Group in Computational Science and HPC, Dhirubhai Ambani University (formerly DA-IICT), Gandhinagar, INDIA. \texttt{jagarwal@ipr.res.in and bhaskar\_chaudhury@dau.ac.in}}

\affil{$^2$Institute for Plasma Researach, Gandhinagar, India - 382428}

\date{ }
\providecommand{\keywords}[1]{\textbf{\textit{Keywords-}} #1}
\begin{document}
\maketitle


\begin{abstract}
Reliable early disruption prediction is critical for the safe operation and real-time control
of tokamaks. However, machine learning based prediction frameworks have
predominantly targeted medium and long pulse devices, with comparatively limited
attention given to short pulse tokamaks where available warning time is inherently
constrained. In this work, an interpretable machine learning framework is developed for feature engineering and early
prediction of disruptions in the ADITYA using the initial
plasma evolution information, prior to the activation of the negative converter of the
ohmic transformer power supply. Statistical descriptors comprising the mean,
variance, skewness, kurtosis and wavelet energy entropy are extracted from routinely
available plasma diagnostics over different operation time windows. Decision tree
based feature selection is employed to identify physically meaningful disruption precursors
and to reduce feature dimensionality. These selected features are used to train a random
forest classifier. The proposed framework achieves stable predictive performance across
different analysis windows, with a maximum ROC-AUC of 0.87 for 0-35 ms and 0-40
ms windows. Comparable and in some cases improved, performance is obtained using the
reduced feature set, demonstrating that the selected statistical descriptors retain the
essential information required for disruption prediction. The proposed methodology
provides an interpretable and computationally efficient framework for real time disruption
prediction in short pulse tokamaks and establishes that carefully engineered statistical
descriptors can effectively replace raw time series inputs for early disruption prediction,
thereby offering a practical pathway toward real time plasma control in short pulse
tokamaks similar to ADITYA and ADITYA-U.
\end{abstract}
\section{Introduction}
The realization of fusion energy through magnetic confinement relies on the stable operation of high temperature plasmas over extended periods. Among the various challenges encountered during tokamak operation, plasma disruptions \cite{schuller1995disruptions, riccardo2003disruptions, boozer2012disruptions} remain one of the most critical obstacles because of their potential to terminate plasma discharges abruptly and generate intense thermal, electromagnetic and mechanical loads on plasma facing components and in vessel structures \cite{mcgrath1990thermal,hollmann2011plasma}. In present day experimental devices, disruptions reduce operational efficiency and experimental availability, whereas in next generation fusion reactors such as ITER and DEMO they are expected to pose a significant threat to machine integrity and component lifetime. Consequently, reliable disruption prediction and mitigation have become indispensable elements of modern tokamak operation and are recognized as key requirements for achieving sustained plasma performance and reactor scale operation.
A plasma disruption is generally preceded by a sequence of nonlinear magnetohydrodynamic (MHD) instabilities \cite{troyon1984mhd, de2008mhd, zohm2003mhd}, impurity accumulation, radiative cooling, current profile modification and confinement degradation, which eventually drive the plasma away from its stable operating regime \cite{koslowski2012operational, zohm2003mhd, jahns1978internal, riccardo2005timescale}. These processes manifest themselves through measurable changes in routinely acquired diagnostic signals, including plasma current, loop voltage, magnetic fluctuations, soft X-ray emission, $H\alpha$ radiation, hard X-ray emission and bolometric radiation. Accurate identification of these precursor signatures sufficiently before the thermal and current quench is essential for enabling active control strategies, such as massive gas injection, shattered pellet injection, or machine specific protective actions, thereby minimizing disruption induced damage and improving operational reliability \cite{jardin2000fast,hollmann2015status, taylor1999disruption, herfindal2019injection, tanna2015mitigation}.
Over the past two decades, disruption prediction has evolved from empirical threshold based approaches to sophisticated data driven techniques capable of exploiting the large volumes of diagnostic data routinely generated by tokamak experiments. Recent advances in machine learning (ML) have significantly improved predictive capability by learning complex nonlinear relationships among multiple plasma diagnostics without requiring explicit physics based models. A broad spectrum of supervised learning algorithms, including artificial neural networks, support vector machines, decision trees (DT), random forest (RF), gradient boosting methods, convolutional neural networks, recurrent neural networks, long short term memory (LSTM) networks, transformers and hybrid deep learning architectures, has been investigated for disruption prediction on several major tokamaks, including JET \cite{aymerich2021statistical,aymerich2022convolutional, aymerich2023comparison,aymerich2024self, croonen2023investigation, pau2019machine}, DIII-D \cite{churchill2020convolutional, fu2020machine, montes2021semi,rea2019predictor}, EAST\cite{deng2025interpretability,guo2021prediction,guo2023disruption}, ASDEX Upgrade\cite{aledda2015improvements}, J-TEXT \cite{wei2019disruption, zheng2020disruption,zheng2023disruption}, KSTAR \cite{lee2025machine}], Alcator C-Mod \cite{montes2019machine}, HL-2A \cite{zhong2021disruption} and HL-3 \cite{yang2025implementing}.  Many studies used the data from multiple tokamaks \cite{spangher2023transformers,kates2019predicting,rea2018disruption,spangher2025disruptionbench,zhu2021adaptive,zhu2021hybrid,zhu2023integrated,zhong2021disruption}. These studies have demonstrated that ML based approaches can achieve high predictive accuracy while providing sufficiently early warning for disruption mitigation systems. Despite these encouraging developments, existing ML frameworks have predominantly been developed and validated for medium  and long pulse tokamaks, where plasma discharges typically extend from several hundred milliseconds to several seconds. Long discharge duration allows models to exploit a substantial temporal evolution of diagnostic signals before disruption onset, enabling the use of sequence based deep learning architectures and large temporal observation windows. However, comparatively, limited attention has been devoted to disruption prediction in short pulse tokamaks, where only limited diagnostic information is available before a disruptive event. Developing reliable prediction models under these conditions remains considerably more challenging compared to long pulse tokamaks.
The ADITYA \cite{bhatt1989aditya, tanna2017overview, bora2002sst} tokamak occupies a unique position among existing disruption prediction studies. Unlike the large superconducting tokamaks on which most ML models have been developed, ADITYA is a short pulse device with a typical plasma discharge duration of approximately 120 ms, while its upgraded successor, ADITYA-U \cite{tanna2018plasma,tanna2019overview, tanna2022overview, tanna2024overview}, extends the discharge duration to nearly 300 ms. The limited plasma duration considerably restricts the amount of diagnostic information available prior to a disruption, thereby reducing the time available for reliable prediction. Consequently, methodologies developed for long pulse tokamaks cannot be directly transferred to short pulse machines without carefully accounting for the reduced temporal evolution of plasma parameters. However, artificial neural networks (ANNs) \cite{sengupta2000forecasting} and LSTM models \cite{agarwal2021sequence} have been explored to predict disruptions in the ADITYA tokamak. Although these efforts have demonstrated the potential of data driven approaches, they have been limited by small and biased datasets which restrict the generalizability and robustness of the models. Recent studies have utilized larger and more diverse datasets from ADITYA and ADITYA-U \cite{joshi2024assessment, muruganandham2024ensemble}, but focused on classifying disruption scenarios rather than providing early predictions. The most recent study \cite{agarwal2026early}, used the transformer based deep learning model  to predict the disruption. While the transformer based approach demonstrated disruption prediction effectively with a maximum warning time of 8 ms, the interpretability of the results remains an area of concern.
An additional operational challenge specific to ADITYA arises from the protection requirements of the OTPS (ohmic transformer power supply). During disruptive discharges, the negative converter stage of the OTPS  needs to be deactivated to avoid unnecessary electrical and thermal stresses on the OTPS components, which otherwise is used to extend the plasma discharge \cite{kumawat2021electrical, balakrishnan2003plasma}. For an effective disruption mitigation strategy, however, the impending disruption must be identified sufficiently early so that appropriate control actions can be initiated before the activation of the negative converter. This requirement places stringent constraints on prediction latency and necessitates reliable disruption prediction using only the initial phase of plasma evolution. Developing models capable of making accurate predictions within approximately the first 20-50 ms of plasma initiation, well before the negative converter is engaged, therefore represents an operationally significant yet largely unexplored problem.\\
 Although deep learning (DL) based approaches have substantially improved disruption prediction performance and achieve excellent predictive accuracy, they often require large volumes of training data, extensive computational resources and long temporal sequences to effectively capture the evolution of plasma dynamics \cite{vega2022disruption, rea2018disruption, guo2021disruption, lee2025machine}. More importantly, the prediction mechanism of these models is frequently difficult to interpret, limiting their ability to provide physical insight into the diagnostic signatures responsible for disruption onset. As ML becomes increasingly integrated into plasma control systems, interpretability is emerging as an important consideration, enabling greater confidence in model predictions while facilitating the identification of physically meaningful disruption precursors.
Motivated by these challenges, this work develops an interpretable ML framework \cite{rea2020progress, deng2025interpretability} for early disruption prediction in the ADITYA tokamak using routinely used plasma diagnostics. Instead of directly employing high dimensional diagnostic time series, each plasma discharge is represented through a set of statistical descriptors that capture both the global characteristics and local fluctuations of the measured signals. Mean, variance, skewness, kurtosis and wavelet energy entropy are extracted from multiple plasma diagnostics over different early discharge observation windows to characterize the evolving plasma behaviour while significantly reducing the dimensionality of the original data. DT \cite{quinlan1986induction, quinlan2014c4, mienye2024survey} based feature selection is subsequently employed to identify the most informative disruption precursors and the resulting reduced feature space is used to train a RF \cite{biau2016random, salman2024random, breiman2001random, cutler2012random} classifier for disruption prediction.
The proposed framework is designed not only to achieve reliable predictive performance but also to improve the physical interpretability of the prediction process. By identifying the diagnostic signals and statistical descriptors that contribute most strongly to classification, the methodology provides insight into the evolution of disruption precursors during the earliest stages of plasma formation. Furthermore, because the framework relies exclusively on routinely available diagnostics and computationally inexpensive statistical features, it is well suited for future real time implementation. Although the present study focuses on ADITYA, the proposed methodology is equally relevant to ADITYA-U and other short pulse tokamaks, where reliable disruption prediction must be achieved using limited temporal information during the initial phase of plasma evolution.
By leveraging only the earliest phases of plasma evolution for disruption prediction in short-pulse tokamaks, this work advances interpretable ML beyond the long-pulse operating regimes prevalent in current literature and establishes a physics-informed pathway toward real-time disruption forecasting.
\section{Experimental dataset and methodology}
\subsection{\label{sec:level1}ADITYA tokamak}
The experimental data used in the present study have been obtained from the ADITYA tokamak operated at the Institute for Plasma Research (IPR), Gandhinagar, India. ADITYA is a medium sized tokamak \cite{bhatt1989aditya} with a major radius of 0.75 m and a minor radius of 0.25 m, designed to investigate plasma confinement, MHD stability, plasma wall interactions and disruption physics under short pulse operating conditions \cite{tanna2019overview, tanna2017overview, bhatt1989aditya}. The device typically operates with hydrogen plasmas at toroidal magnetic fields of up to 1.2 T and plasma currents in the range of 80-250 kA, producing discharge durations of approximately 120 ms.
Although the relatively short pulse length of ADITYA limits the total plasma duration compared with large superconducting tokamaks, it provides a unique experimental platform for investigating disruption prediction under stringent temporal constraints. 
\subsection{\label{sec:level1}ADITYA diagnostics}
The proposed disruption prediction framework utilizes six routinely measured plasma diagnostic signals available during ADITYA operation. These diagnostics have been selected because they collectively describe the evolution of plasma equilibrium, current dynamics, impurity behaviour, energetic particle activity and radiative losses \cite{bhatt1989aditya, tanna2017overview, raju2000mirnov, atrey2018mhd,bora2002sst}.
The plasma current ($I_p$) represents the global evolution of the discharge and is one of the primary indicators of plasma stability. Variations in the current waveform often precede disruption onset and therefore provide valuable information regarding the evolving plasma state. The loop voltage ($V_{\mathrm{loop}}$) reflects the electrical conditions required to sustain the plasma current and is sensitive to changes in plasma resistivity during discharge evolution. Optical diagnostics are represented by the H$\alpha$ emission signal, which characterizes plasma wall interactions and neutral particle recycling. Enhanced H$\alpha$ radiation frequently accompanies increased impurity influx and deteriorating confinement conditions, making it an important indicator of disruption related plasma behaviour. Radiative diagnostics include soft X-ray (SXR) emission and bolometric radiation (Bolo), both of which provide complementary information regarding plasma energy losses. While SXR measurements primarily monitor radiation originating from the hot plasma core, bolometric measurements represent the total radiated power and therefore provide a global assessment of impurity radiation and plasma cooling. Hard X-ray (HXR) emission is included to characterize energetic and runaway electron populations that may develop during the evolution of disruptive discharges. Changes in HXR activity often accompany modifications in plasma confinement and energy transport, providing additional information regarding the progression toward disruption \cite{agarwal2024labelling,agarwal2026early}.
The combination of these six diagnostics provides complementary observations of plasma behaviour from multiple physical perspectives. Since all measurements are routinely available during normal ADITYA operation, the proposed methodology does not require specialized diagnostics or additional instrumentation, thereby enhancing its practical applicability for future real time implementation.
\subsection{\label{sec:level1}Experimental database and data curation}
The experimental database used in this study consists of plasma discharges acquired during routine operation of the ADITYA tokamak. The dataset includes both disruptive and non disruptive plasma discharges, enabling the disruption prediction problem to be formulated as a supervised binary classification task. Each discharge was assigned a class label according to its terminal plasma behaviour, where disruptive discharges correspond to plasmas terminated by a disruption and non disruptive discharges correspond to plasmas that completed the discharge without exhibiting disruptive behaviour. The final database comprises 541 disruptive and 637 non disruptive plasma discharges, resulting in a total of 1178 labelled discharges used for model development and performance evaluation.

To investigate the earliest stage at which reliable prediction can be achieved, seven observation windows have been considered during the initial plasma evolution. Diagnostic signals have been analysed from plasma initiation up to 20, 25, 30, 35, 40, 45 and 50 ms, respectively. These observation windows have been deliberately selected to evaluate the trade off between prediction latency and prediction accuracy. Shorter observation windows provide earlier warning of impending disruptions but contain comparatively less information regarding the evolving plasma dynamics. Conversely, larger observation windows capture additional disruption precursors but reduce the available response time for mitigation. Evaluating multiple observation windows therefore enables identification of the earliest interval that provides reliable predictive capability while preserving sufficient time for future control actions.
Prior to feature extraction, all diagnostic signals have been synchronized with respect to plasma initiation and examined for consistency across the complete dataset. 

\subsection{\label{sec:level1}Overview of the proposed methodology}
\begin{figure}[h]
    \centering
    \includegraphics[width=\linewidth]{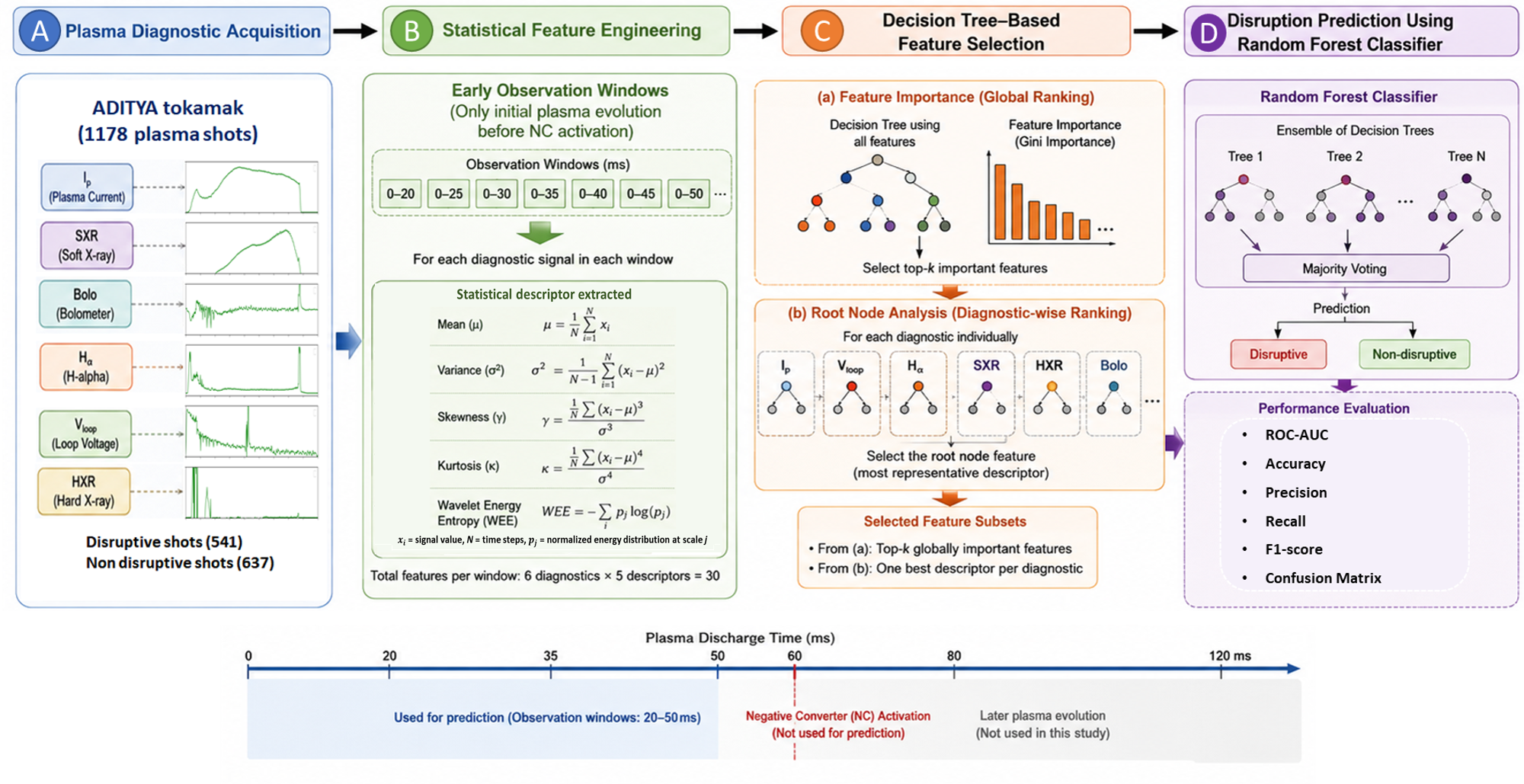}
    \caption{Schematic overview of the proposed interpretable ML framework for early disruption prediction in the ADITYA tokamak}
    \label{fig:overview}
\end{figure}

A schematic overview of the proposed disruption prediction framework is presented in Figure \ref{fig:overview}. The methodology consists of four sequential stages: (i) acquisition of plasma diagnostic signals, (ii) statistical feature extraction, (iii) DT based feature selection and (iv) disruption prediction using a RF classifier.
In the first stage, the six plasma diagnostic signals described in section 2.2 and 2.3 are acquired over each of the predefined observation windows. Rather than directly employing the raw time series data for classification, each diagnostic signal is represented using a set of statistical descriptors that characterize its temporal behaviour (block B in Figure \ref{fig:overview}). This representation substantially reduces the dimensionality of the original dataset while preserving the dominant statistical characteristics associated with plasma evolution.
The extracted statistical descriptors are subsequently analysed using two complementary DT based feature selection strategies (block C in Figure \ref{fig:overview}). In the first approach, feature importance values obtained from the complete DT model are used to identify the most informative statistical parameters. In the second approach, individual diagnostic signals are analysed independently and the statistical descriptor selected as the root node of each DT is considered the most representative feature for that diagnostic. These complementary strategies enable identification of a compact subset of statistically and physically relevant disruption precursors while reducing redundancy within the feature space.
Finally, the selected features are supplied to a RF classifier for disruption prediction (block D in Figure \ref{fig:overview}). The classifier is evaluated using both the complete feature space and the reduced feature sets obtained from the DT analysis, allowing the influence of feature selection on predictive performance to be systematically assessed. The overall workflow therefore provides a unified framework for evaluating the effectiveness of interpretable statistical feature engineering in early disruption predictions.
\section{Statistical feature engineering}
As described in the previous section, rather than analysing the complete time series directly, the proposed approach summarizes the temporal behaviour of each diagnostic signal using a small number of physically interpretable statistical parameters. 
For each diagnostic signal, five complementary statistical descriptors have been calculated, mean, variance, skewness, kurtosis and wavelet energy entropy. 
The mean represents the average value of a diagnostic signal within the selected observation window and provides a measure of its overall operating level. Although the mean alone cannot characterize transient fluctuations, it reflects the general evolution of plasma parameters during discharge initiation and therefore serves as a useful baseline descriptor for comparing disruptive and non disruptive plasmas. The variance quantifies the dispersion of a diagnostic signal about its mean value and therefore provides information regarding the magnitude of temporal fluctuations during plasma evolution. Since disruption precursors are frequently accompanied by increasing instability and enhanced fluctuations in several plasma diagnostics, the variance can serve as an indicator of the evolving disruption dynamics. Larger variance values generally indicate greater departures from steady plasma behaviour and may therefore contain valuable information regarding the onset of instability.
While the mean and variance describe the central tendency and dispersion of a diagnostic signal, they do not characterize the symmetry of its statistical distribution. Skewness  quantifies the degree of asymmetry in the signal distribution about its mean. During stable plasma operation, many diagnostic signals exhibit relatively symmetric fluctuations, whereas the development of disruption precursors often introduces intermittent excursions associated with growing MHD activity, impurity influx, or enhanced radiation losses. Consequently, skewness provides additional information regarding the evolving plasma dynamics that cannot be captured by mean and variance alone.Positive and negative skewness values indicate asymmetric distributions dominated by large positive or negative excursions, respectively, while values close to zero correspond to nearly symmetric signal distributions. Kurtosis measures the concentration of data around the mean together with the presence of extreme deviations. Unlike variance, which quantifies the overall spread of a signal, kurtosis is particularly sensitive to infrequent but large amplitude fluctuations. Such transient events frequently accompany the nonlinear evolution of plasma instabilities preceding disruption and therefore provide valuable information regarding disruption precursor dynamics. Higher kurtosis values generally indicate the occurrence of intermittent bursts or sharp transient events, whereas lower values correspond to comparatively uniform signal fluctuations. Since disruption precursors frequently manifest as localized bursts in plasma diagnostics, kurtosis provides an effective statistical measure for distinguishing disruptive from non disruptive plasma behaviour.

Although the statistical descriptors discussed above effectively characterize the amplitude distribution of plasma diagnostic signals, they do not explicitly represent the temporal evolution of signal energy across different frequency scales. Plasma disruptions are inherently multiscale phenomena involving the simultaneous interaction of MHD instabilities, impurity radiation, energetic particles and turbulence, all of which evolve over different temporal scales. Consequently, a feature capable of quantifying the complexity of these multiscale dynamics is desirable for early disruption prediction. For the present study, the discrete wavelet transform was implemented using the Daubechies 4 (db4) wavelet. The signal was decomposed up to four levels, or the maximum permissible level for shorter signals, whichever was smaller. The db4 wavelet was selected because of its good time frequency localization properties and its suitability for analyzing non stationary plasma signals. The signal $x(t)$ is decomposed into multiple scales using discrete wavelet transform, yielding wavelet coefficients. Then entropy is evaluated using the normalized energies of the approximation and detail coefficients obtained from the wavelet decomposition.
Lower entropy values indicate that the signal energy is concentrated within a limited number of frequency bands, whereas higher entropy values correspond to a broader distribution of energy across multiple scales, reflecting increased signal complexity. Since the onset of plasma disruption is frequently accompanied by the emergence of multiscale fluctuations and evolving instability dynamics, wavelet energy entropy provides an effective descriptor of the changing complexity of plasma behavior during the initial phase of discharge evolution.

Collectively, these descriptors characterize the average behaviour, variability, distribution asymmetry, fluctuations and multiscale complexity of the evolving plasma signals. The resulting feature vector therefore captures both the global statistical properties and localized transient behaviour associated with disruption precursor dynamics.

\section{Decision tree based feature selection}
The statistical feature engineering framework described in the previous section generates a comprehensive set of descriptors representing multiple plasma diagnostics across different observation windows. Set of descriptors for single observation window is given in table \ref{tab:features}. Although this feature representation captures diverse aspects of plasma behaviour, not all statistical descriptors contribute equally to disruption prediction. The presence of redundant or weakly informative features may increase computational complexity, reduce model interpretability and potentially degrade the generalization capability of the classifier. Consequently, an effective feature selection strategy is required to identify the most informative disruption precursors while preserving the essential predictive information.

\begin{table}[!h]
\centering
\caption{Statistical descriptor passed as features for specific observation window}
\label{tab:features}
\small
\begin{tabular}{ll}
\hline
\textbf{Statistical descriptor} &
\textbf{Nomenclature} \\
\hline
Variance in plasma current & $Ip_v$ \\

Mean of plasma current & $Ip_m$   \\

Skewness of plasma current & $Ip_s$   \\

Kurtosis of plasma current & $Ip_k$   \\

Wavelet energy entropy of plasma current & $Ip_{wee}$   \\

Variance in bolometer & $Bolo_v$ \\

Mean of bolometer & $Bolo_m$   \\

Skewness of bolometer & $Bolo_s$   \\

Kurtosis of bolometer & $Bolo_k$   \\

Wavelet energy entropy of bolometer & $Bolo_{wee}$   \\

Variance in H$\alpha$ & $H\alpha_v$ \\

Mean of H$\alpha$ & $H\alpha_m$   \\

Skewness of H$\alpha$ & $H\alpha_s$   \\

Kurtosis of H$\alpha$ & $H\alpha_k$   \\

Wavelet energy entropy of H$\alpha$ & $H\alpha_{wee}$   \\

Variance in HXR & $HXR_v$ \\

Mean of HXR & $HXR_m$   \\

Skewness of HXR & $HXR_s$  \\

Kurtosis of HXR & $HXR_k$   \\

Wavelet energy entropy of HXR & $HXR_{wee}$   \\

Variance in SXR & $SXR_v$ \\

Mean of SXR & $SXR_m$   \\

Skewness of SXR & $SXR_s$   \\

Kurtosis of SXR & $SXR_k$   \\

Wavelet energy entropy of SXR & $SXR_{wee}$   \\

Variance in C111 & $C111_v$ \\

Mean of C111 & $C111_m$   \\

Skewness of C111 & $C111_s$   \\

Kurtosis of C111 & $C111_k$   \\

Wavelet energy entropy of C111 & $C111_{wee}$   \\
\hline
\end{tabular}
\end{table}
DTs provide an attractive framework for this purpose because they combine classification capability with intrinsic feature interpretability \cite{lungaroni2018potential, angelino2018learning, ribeiro2016should}. Unlike many feature selection techniques that rank variables solely on statistical criteria, a DT identifies the sequence of features that most effectively separates disruptive and non disruptive plasma discharges through recursive binary partitioning of the feature space. Each internal node represents a decision based on a particular statistical descriptor, while the terminal nodes correspond to the predicted plasma class. Since the features appearing near the root of the tree contribute most strongly to the classification process, the resulting tree structure provides direct insight into the relative importance of different plasma diagnostics and their associated statistical characteristics.
In the present study, DTs have been employed exclusively as a feature selection tool rather than as the final disruption predictor. This distinction is important because the objective is not to maximize the standalone predictive performance of a single DT but to identify a compact and physically meaningful subset of statistical descriptors that can subsequently be used to train a more robust ensemble classifier. The selected features therefore represent the dominant disruption signatures extracted from the statistical feature space and provide the foundation for the RF classifier described in the subsequent section.
To obtain a comprehensive assessment of feature relevance, two complementary feature selection strategies have been investigated. The first evaluates the importance of statistical descriptors across the complete feature space, thereby identifying the plasma diagnostics that contribute most significantly to disruption prediction when all information is considered simultaneously. The second examines each diagnostic independently, allowing the most representative statistical descriptor to be identified for every plasma measurement. Together, these complementary analyses provide both a global and diagnostic specific understanding of disruption precursor evolution.
\subsection{\label{sec:level1}Decision tree algorithm}
\begin{figure}[h]
\centering
\includegraphics[width=0.8\textwidth]{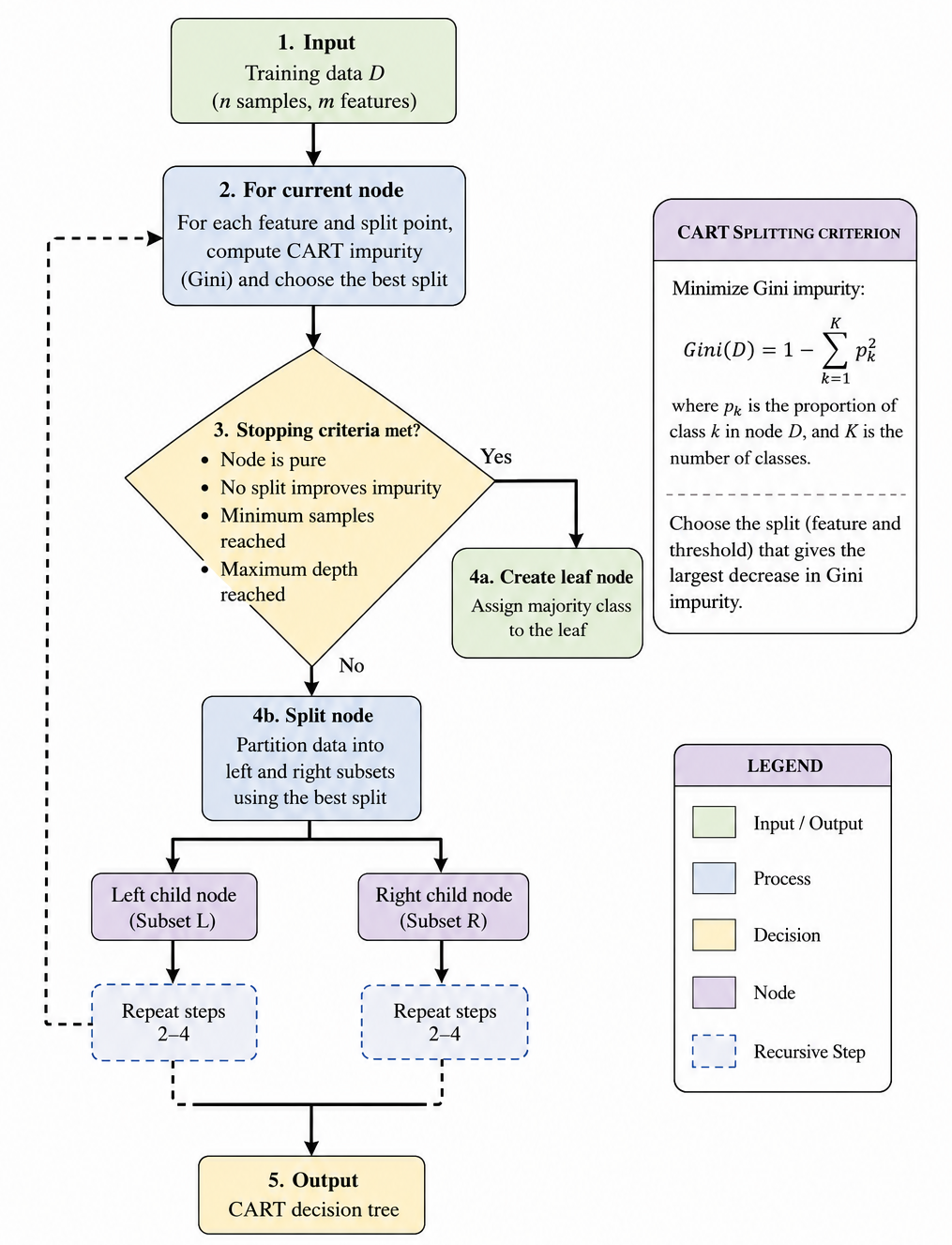}
\caption{Flowchart of the CART decision tree classifier}
\label{fig:tree}
\end{figure}

A DT classifier based on the classification and regression tree (CART) algorithm  \cite{breiman2017classification} is adopted for feature selection because of their transparent decision making structure and suitability for feature driven classification. The CART algorithm constructs a binary DT by recursively selecting the feature and corresponding threshold that maximize the reduction in node impurity. A schematic representation of this algorithm is shown in Fig.  \ref{fig:tree}. 
As illustrated, the process begins at the root node, where the optimal feature is selected as the splitting criterion. The dataset is then partitioned into branches according to the possible outcomes of the attribute test. This recursive procedure continues for subsequent nodes until a stopping condition is satisfied, resulting in leaf nodes that represent the final class labels (e.g., disruptive or non disruptive states). Let the training dataset at a given node be denoted by $S$ which contains samples from two classes, disruptive ($y=1$) and non disruptive ($y=0$). )). The impurity of node (S) is quantified using the gini impurity, defined as given in Eq \ref{eq:gini}.
\begin{equation}
G(S) = 1 - \sum_{c \in \{0,1\}} p(c)^2
\label{eq:gini}
\end{equation}
Where $p(c)$ is the proportion of samples in $S$ that belong to class $c$. The Gini impurity is zero when all samples belong to a single class and reaches its maximum value when the classes are equally mixed.

For a candidate split based on feature $X$ and threshold $t$, the dataset $S$ is partitioned into two subsets, $S_L$ and $S_R$, containing samples that satisfy $X \leq t$ and $X > t$, respectively. The resulting impurity after the split is given by Eq. \ref{eq:impurity}.
\begin{equation}
G_{\text{split}}(S,X,t)
\frac{|S_L|}{|S|}G(S_L)
+
\frac{|S_R|}{|S|}G(S_R)
\label{eq:impurity}
\end{equation}
Where $|S_L|$ and $S_R$ denotes the number of samples in left and right subsets, respectively.
The reduction in impurity achieved by the split as shown in Eq. \ref{eq:split}.

\begin{equation}
\Delta G(S,X,t)
G(S)-G_{\text{split}}(S,X,t)
\label{eq:split}
\end{equation}

At each node, the CART algorithm selects the feature and threshold combination that maximizes the impurity reduction, $\Delta G(S,X,t)$, thereby yielding the greatest improvement in class separation. This greedy top down procedure is applied recursively until a predefined stopping criterion such as the maximum tree depth or a pure leaf node, is reached to identify the most informative diagnostic features. 

\subsection{\label{sec:level1}Multi diagnostic feature importance analysis}
In the first feature selection strategy, the complete statistical feature matrix obtained from all six plasma diagnostics was supplied to the DT algorithm. The resulting tree was analysed using the feature importance coefficients, which quantify the contribution of each statistical descriptor to the reduction of classification impurity throughout the tree construction process. Features associated with larger importance values contribute more substantially to the discrimination between disruptive and non disruptive plasma discharges and are therefore considered stronger disruption precursors. Here the feature which has importance value more than 0.01 is considered.
The analysis was performed independently for each observation window between 20 and 50 ms to investigate the temporal evolution of feature relevance during the early stages of plasma formation. Since the available diagnostic information increases with observation window length, the dominant statistical descriptors are expected to evolve as the plasma progresses from discharge initiation toward disruption onset. This time resolved feature selection therefore provides valuable physical insight into the changing diagnostic signatures associated with disruption development.
\subsection{\label{sec:level1}Diagnostic specific feature selection}
While the global feature importance analysis identifies the most informative statistical descriptors across the complete feature space, it does not reveal which statistical characteristic is most representative for an individual plasma diagnostic. Since each diagnostic monitors a different physical aspect of plasma behaviour, the statistical descriptor that best characterizes disruption precursors may vary from one diagnostic to another. To address this issue, a second feature selection strategy was adopted in which each diagnostic signal was analysed independently using a separate DT.
For every diagnostic, the corresponding statistical descriptors (mean, variance, skewness, kurtosis and wavelet energy entropy) have been supplied to the DT algorithm and the descriptor selected at the root node was identified as the most discriminative feature for that diagnostic. Because the root node performs the first partition of the feature space, it represents the statistical characteristic that provides the greatest reduction in class impurity and therefore carries the highest discriminative power for distinguishing disruptive from non disruptive plasma discharges.
The analysis was repeated for each observation window between 20 and 50 ms to investigate how the dominant statistical descriptor evolves during the initial phase of plasma formation. Unlike the multi diagnostic feature importance analysis, which compares different diagnostics simultaneously, this approach isolates the statistical behaviour of each individual plasma measurement. Consequently, it provides a clearer understanding of how the information content of each diagnostic evolves as disruption precursors develop.

\section{\label{sec:level1}Random forest based disruption prediction}
Following the DT based feature selection described in the previous section, disruption prediction was performed using a RF classifier. RF was selected because it combines the predictions of multiple DTs to produce a robust ensemble model that generally exhibits improved classification accuracy, reduced overfitting and better generalization compared with a single DT \cite{becker2023decision}. In addition, RF can effectively model complex nonlinear relationships among plasma diagnostic features while maintaining computational efficiency, making it well suited for disruption prediction problems.
In the proposed framework, the RF classifier constitutes the final stage of the prediction pipeline. Statistical descriptors extracted from the six plasma diagnostic signals are used as the input feature vector. Two feature representations have been investigated. In the first, the complete statistical feature set generated in Section 3 was supplied directly to the classifier. In the second, only the reduced feature subsets identified using the DT based feature selection strategy have been used. This comparison enables evaluation of whether a compact, physically interpretable feature representation can preserve the predictive information contained in the complete feature space.

\subsection{\label{sec:level1}Random forest algorithm}
RF is an ensemble learning method based on bootstrap aggregation (bagging). Given a training dataset of size $N$, multiple bootstrap samples are generated by sampling with replacement from the original dataset. For each bootstrap sample, an independent DT $h_t(\mathbf{X})$ is trained. Furthermore, at each split in the tree construction process, a random subset of features $m \subset d$ (where $d$ is the total number of features) is considered. This randomization reduces correlation among individual trees and improves ensemble robustness.

The final prediction $\hat{y}$ of the RF is obtained through majority voting shown in Eq. \ref{eq:voting}.
\begin{equation}
\hat{y} = \mathrm{mode} \{ h_1(\mathbf{X}), h_2(\mathbf{X}), \dots, h_T(\mathbf{X}) \} 
\label{eq:voting}
\end{equation}

where $T = 200$ denotes the total number of trees (corresponding to \texttt{n\_estimators = 200}). 

The RF classifier was implemented using the scikit learn ML library. Hyperparameters that are not explicitly optimized are assigned their default values. The principal model parameters are summarized in table \ref{tab:rf_parameters}.

\begin{table}[!h]
\caption{Random forest hyperparameters used in the present study.}
\label{tab:rf_parameters}
\centering
\begin{tabular}{ll}
\hline
\textbf{Parameter} & \textbf{Value} \\
\hline
\texttt{n\_estimators} & 200 \\
\texttt{max\_depth} & None \\
\texttt{max\_features} & sqrt \\
\texttt{min\_samples\_split} & 2 \\
\texttt{min\_samples\_leaf} & 1 \\
\hline
\end{tabular}
\end{table}

Here \texttt{n\_estimators} specifies the number of trees in the ensemble, \texttt{max\_depth} controls the maximum depth of individual trees, while \texttt{max\_features} determines the number of features randomly considered at each split. The \texttt{min\_samples\_split} and \texttt{min\_samples\_leaf} define the minimum number of samples required for node splitting and leaf node formation, respectively. Bootstrap aggregation was enabled to promote diversity among individual trees. To mitigate the effect of class imbalance, class weights have been adjusted automatically. A fixed random seed was used to ensure reproducibility. Unlike the single DT model, no explicit restriction was imposed on tree depth ($\texttt{max\_depth} = \texttt{None}$), allowing individual trees to grow fully and capture complex nonlinear interactions among the extracted statistical features.
The variance reduction property of the ensemble can be approximated as given in Eq. \ref{eq:reduction}.

\begin{equation}
  \mathrm{Var}(\hat{y}) \approx \rho \sigma^2 + \frac{1 - \rho}{T} \sigma^2  
  \label{eq:reduction}
\end{equation}

where $\rho$ represents the correlation between individual trees and $\sigma^2$ denotes the variance of a single tree. This expression illustrates that, as the number of trees $T$ increases, the overall variance of the ensemble decreases, thereby improving model stability and generalization performance. 
For this study the dataset was partitioned into training and testing subsets separately for disruptive and non disruptive discharges to preserve the class distribution in both sets. The primary objective of the present study is to identify physically interpretable disruption pathways and influential diagnostic features rather than to optimize benchmark classification performance. A more extensive cross validation analysis could further quantify the statistical robustness of the predictive performance and will be considered in future studies.
\subsection{\label{sec:level1}Model development and performance evaluation}
The disruption prediction problem has been formulated as a supervised binary classification task in which each plasma discharge is assigned to either the disruptive or non disruptive class. The statistical feature vectors generated from the selected observation windows served as the input to the RF classifier, while the corresponding discharge labels constitutes the target output.
To provide an unbiased assessment of classifier performance, the dataset was divided into independent training and testing subsets. Equal numbers of disruptive and non disruptive discharges have been included in the test set to avoid performance bias arising from class imbalance. Here total test samples are 216 experiments in which 108 belong to disruptive class and rest 108 belongs to non disruptive class. The remaining plasma discharges are used for model training. The same data partition was employed for both the complete feature representation and the reduced feature representation obtained from the DT analysis, ensuring a fair comparison between the two approaches.
The RF classifier was trained independently for each observation window between 20 and 50 ms, allowing the predictive capability of the statistical descriptors to be evaluated as a function of the available plasma evolution. This strategy enables systematic investigation of the trade off between prediction latency and classification performance, which is of particular importance for short pulse tokamaks where only a limited interval is available for disruption prediction.

The predictive performance of the RF classifier was assessed using multiple complementary evaluation metrics, including accuracy, precision, recall, F1 score and the area under the receiver operating characteristic curve (ROC-AUC). While accuracy provides an overall measure of classification performance, precision and recall quantify the classifier's ability to correctly identify disruptive discharges while minimizing false alarms and missed disruptions. The F1 score provides a balanced assessment of precision and recall, whereas the ROC-AUC evaluates the discrimination capability of the classifier independently of the selected decision threshold.
In addition to evaluating the complete statistical feature representation, the same performance metrics have been calculated for the reduced feature sets obtained from the DT based feature selection stage. The resulting performance comparison is presented and discussed in the following section.
\section{\label{sec:level1}Results and discussion}
Since the statistical descriptors derived from multiple plasma diagnostics are not expected to contribute equally to disruption prediction, a DT-based feature selection framework was employed to identify the most informative features improving model interpretability and computational efficiency \cite{noroozi2023analyzing}. To evaluate its effectiveness, the predictive performance of the RF classifier is assessed using both the complete and reduced feature sets. The results are presented in two parts. First, the feature selection outcomes are analyzed to identify the most influential disruption precursors. Subsequently, the predictive performance of the RF classifier trained on the complete and reduced feature sets is compared.

\subsection{Decision tree feature selection results}
The feature importance analysis was performed independently for each observation window to identify the statistical descriptors that contribute most significantly to the discrimination between disruptive and non disruptive plasma discharges. The corresponding feature importance distributions are presented in table \ref{tab:feature}, while the diagnostic specific root node selections are summarized in table \ref{tab:rootnode}.
\begin{table*}[!ht]
\centering
\caption{Important parameters with respect to experimental time window}
\label{tab:feature}
\small
\begin{tabular}{ll}
\hline
\textbf{Time window} &
\textbf{Important feature}\\
\hline
0-20 ms & $ Bolo_s,\, Bolo_v,\,C111_k,\, C111_m,\, C111_s,\,  H\alpha_m,\, H\alpha_{wee},\, HXR_m,\, HXR_v,\,Ip_m,\, SXR_m,\, SXR_v $  \\

0-25 ms & $ Bolo_m,\, Bolo_v,\, C111_m,\, HXR_m,\, HXR_v,\,Ip_m,\, Ip_{wee},\,SXR_m $    \\

0-30 ms & $ Bolo_s,\, Bolo_v,\,C111_k,\, C111_m,\, H\alpha_v,\,  HXR_k,\, HXR_{wee},\, HXR_m,\, Ip_k,\,Ip_v,\, SXR_m,\, SXR_v$   \\

0-35 ms & $Bolo_m,\, Bolo_{wee},\, C111_m,\, HXR_k,\, HXR_{wee},\,Ip_m,\,Ip_v,\, Ip_{wee},\,SXR_m$  \\

0-40 ms & $Bolo_m,\, Bolo_v,\, \, Bolo_{wee},\,C111_m,\, HXR_{wee},\, HXR_v,\,Ip_m,\,Ip_v,\, Ip_{wee},\,SXR_m,\, SXR_{wee} $  \\

0-45 ms & $ C111_m,\, C111_s,\,  H\alpha_s,\, HXR_v,\, HXR_{wee},\,Ip_m,\, Ip_v,\,Ip_{wee},\, SXR_{wee} $  \\

0-50 ms & $Bolo_k,\,Bolo_{wee},\, C111_m,\, C111_s,\, Ip_m,\, Ip_v,\, Ip_{wee},\,SXR_v,\, SXR_{wee}$  \\
\hline
\end{tabular}
\end{table*}
\begin{table}[htbp]
\centering
\caption{Root node feature selected by the DT classifier for each diagnostic signal across different analysis windows.}
\label{tab:rootnode}
\small
\renewcommand{\arraystretch}{1.15}

\begin{tabular}{lcccccc}
\toprule
\multirow{2}{*}{\shortstack{\textbf{Time}\\\textbf{window}}} &
\multicolumn{6}{c}{\textbf{Root node feature selected from each diagnostic}} \\
\cmidrule(lr){2-7}
& \textbf{Bolometer} & \textbf{$C$-III} & \textbf{$H\alpha$}
& \textbf{HXR} & \textbf{$I_p$} & \textbf{SXR} \\
\midrule
0--20 ms & $Bolo_v$ & $CIII_m$ & $H\alpha_{wee}$ & $HXR_s$ & $I_{p,s}$ & $SXR_m$ \\
0--25 ms & $Bolo_v$ & $CIII_m$ & $H\alpha_v$ & $HXR_v$ & $I_{p,v}$ & $SXR_m$ \\
0--30 ms & $Bolo_v$ & $CIII_m$ & $H\alpha_v$ & $HXR_v$ & $I_{p,v}$ & $SXR_m$ \\
0--35 ms & $Bolo_s$ & $CIII_m$ & $H\alpha_v$ & $HXR_v$ & $I_{p,v}$ & $SXR_m$ \\
0--40 ms & $Bolo_s$ & $CIII_m$ & $H\alpha_v$ & $HXR_v$ & $I_{p,v}$ & $SXR_m$ \\
0--45 ms & $Bolo_k$ & $CIII_s$ & $H\alpha_{wee}$ & $HXR_v$ & $I_{p,v}$ & $SXR_m$ \\
0--50 ms & $Bolo_k$ & $CIII_k$ & $H\alpha_{wee}$ & $HXR_v$ & $I_{p,wee}$ & $SXR_{wee}$ \\
\bottomrule
\end{tabular}

\vspace{1mm}

\end{table}
The analysis demonstrates that only a relatively small subset of the extracted statistical descriptors dominates the disruption prediction process. Although the complete feature space contains statistical parameters derived from all six plasma diagnostics, the DT consistently identifies a compact group of features that provides the greatest reduction in classification impurity.
A comparison of the feature rankings across different observation windows further reveals that the importance of individual statistical descriptors evolves with plasma development. During the earliest observation windows, first and second order statistical moments dominate the feature rankings, whereas higher order statistical descriptors and wavelet energy entropy become increasingly important as additional plasma evolution is included in the analysis. This behaviour suggests that disruption precursors evolve from relatively simple macroscopic variations during plasma initiation to increasingly complex fluctuation patterns as the plasma approaches disruption.
The diagnostic specific root node analysis provides complementary information regarding the statistical behaviour of individual plasma diagnostics. Rather than comparing all diagnostics simultaneously, this analysis identifies the statistical descriptor that best characterizes each diagnostic independently. The results indicate that different plasma measurements are most effectively represented by different statistical descriptors, reflecting the diverse physical processes associated with plasma current evolution, impurity radiation, energetic electron dynamics and radiative energy losses.
Taken together, the two DT analyses demonstrate that the statistical feature engineering framework successfully identifies a compact and physically meaningful representation of disruption precursor dynamics. These selected features subsequently form the reduced feature set used for RF classification.

\subsection{ Comparison of RF prediction performance for complete and reduced feature set }
\begin{figure}[h]
\centering
\includegraphics[width=1\textwidth]{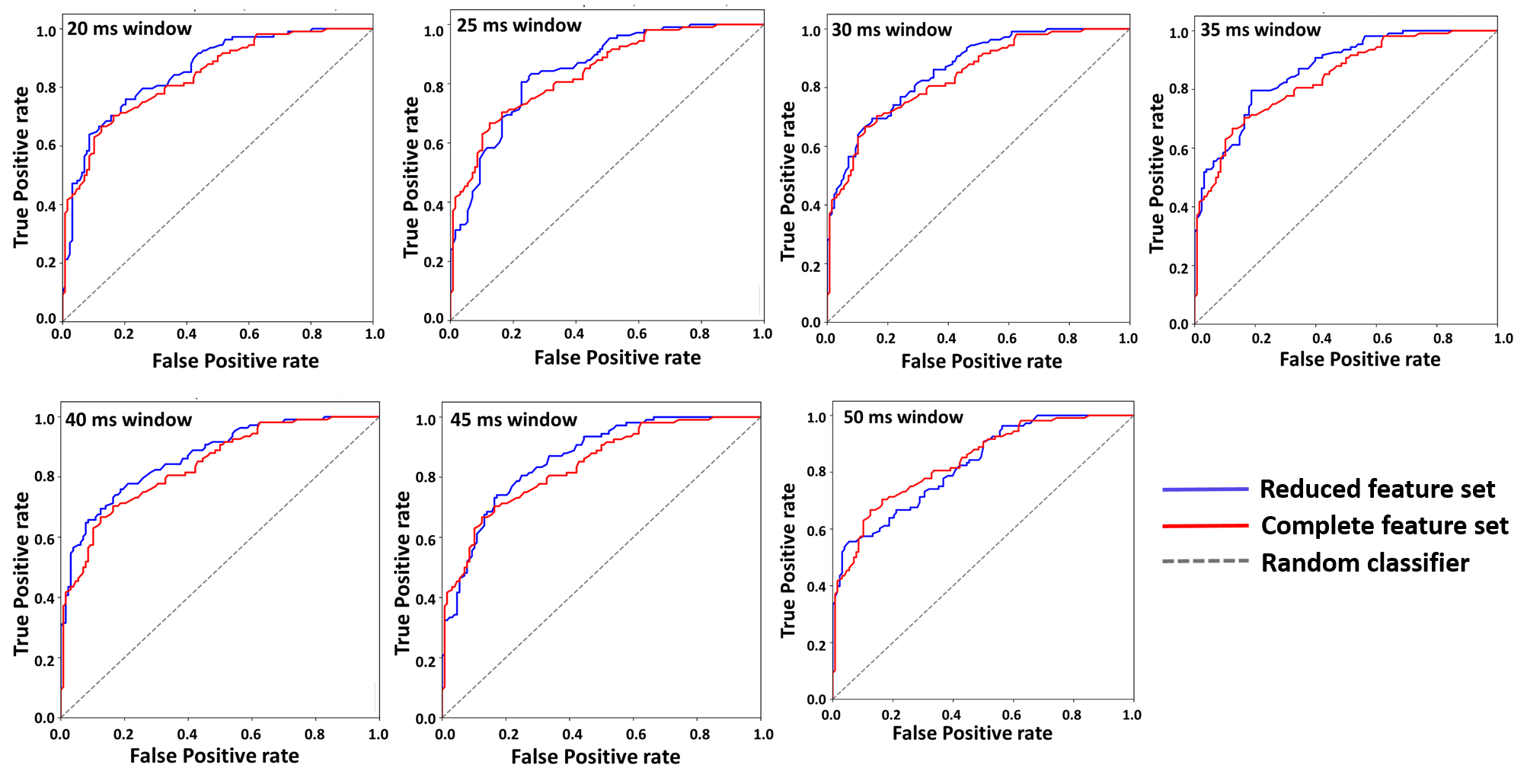}
\caption{ROC curves for different time windows}
\label{fig:ROC}
\end{figure}

\begin{figure}[!ht]
    \centering

        \includegraphics[width=\linewidth]{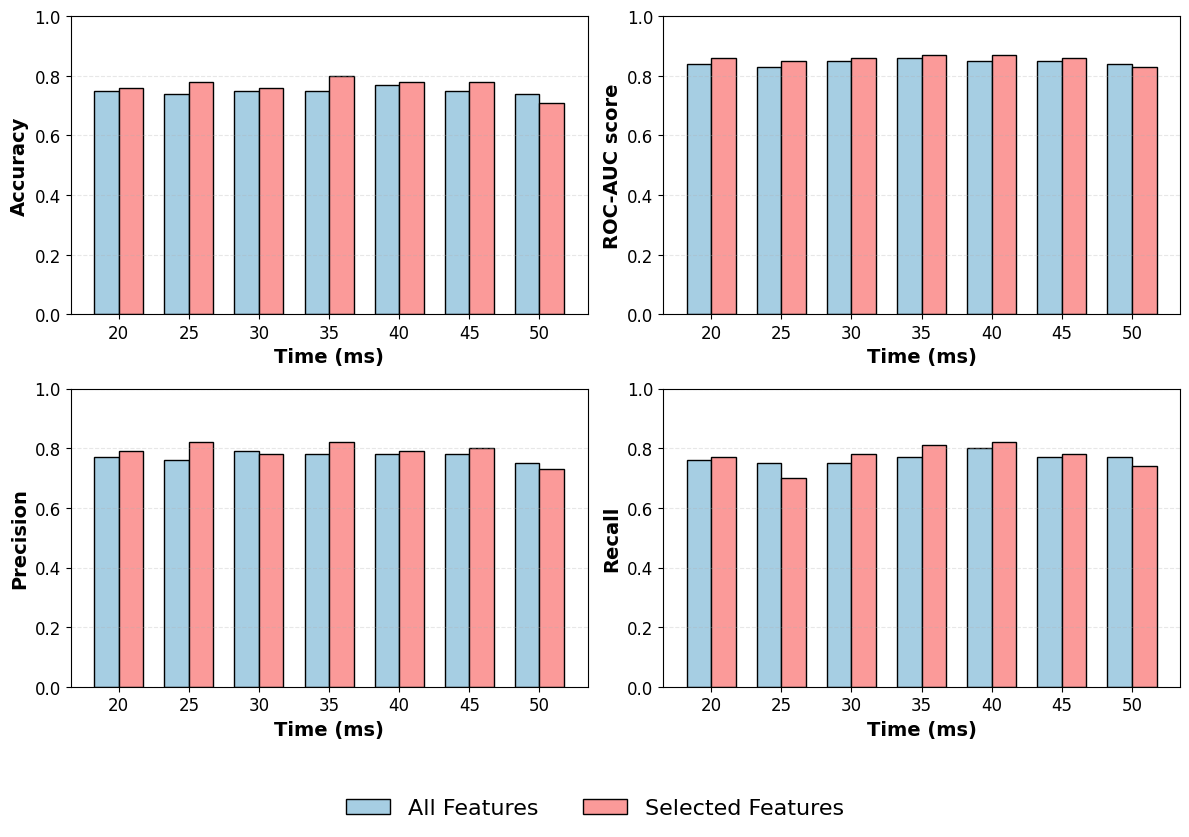}
        \label{fig:RF1_results}
    
    \caption{Variation of the metrics (a) accuracy, (b) ROC-AUC score, (c) precision and (d) recall for RF classifier using complete feature set and reduced feature set with sample time window}
    \label{fig:RF}
\end{figure}

 Classification performance has been assessed independently for each observation window using the evaluation metrics described in Section 5. The receiver operating characteristic (ROC) curves obtained for all observation windows are presented in Fig. \ref{fig:ROC}, while the corresponding performance metrics are shown in Fig. \ref{fig:RF}. The proposed framework demonstrates consistently high predictive performance across the investigated observation windows, confirming that the statistical descriptors extracted during the early phase of plasma evolution contain sufficient information for reliable disruption prediction. While this can also be attributed, in part, to the inherent robustness of RF against redundant and irrelevant features. However, the use of the same classifier for both the complete and reduced feature sets enables a fair comparison.

Among the investigated observation windows, the highest predictive performance is obtained for the 0-35 ms and 0-40 ms observation windows, where the classifier achieves a maximum ROC-AUC of 0.87. Beyond 40 ms, a small decline in predictive performance is observed. A comparison between the complete statistical feature representation and the reduced feature subsets demonstrates that the reduced feature representation consistently achieves predictive performance comparable to that of the complete statistical feature set across all observation windows. In several cases, the reduced feature set yields marginal improvements in ROC-AUC, precision, recall and F1 score, despite containing substantially fewer input variables. These observations indicate that the DT based feature selection successfully removes redundant and weakly informative descriptors while preserving the dominant statistical characteristics associated with disruption precursors.

To further assess the robustness of this observation, 95\% confidence intervals have been estimated for the 40 ms time window using bootstrap resampling of the independent test set. The selected feature model achieved an accuracy of 0.78 with 95\% CI between 0.73-0.83 and an AUC of 0.87 with 95\% CI between 0.82-0.91 in comparison with accuracy of 0.77 with 95\% CI between 0.71-0.82 and AUC of 0.85 with 95\% CI between 0.80-0.89, for the complete feature model. While the selected feature model exhibits marginally higher performance metrics, the confidence intervals demonstrate consistent classifier performance across repeated sampling, indicating that the observed predictive capability is not dependent on a particular train test partition of the dataset. The relatively narrow confidence intervals further confirm the robustness and generalization capability of the proposed disruption prediction framework. To provide additional insight into the classification results at the selected operating threshold, the confusion matrices for the 40 ms prediction window are presented in Table \ref{tab:confusion}, which shows that the reduced feature model yields a slightly better balance of true positives and true negatives while maintaining performance comparable to that of the complete feature set.

\begin{table}[!t]
\caption{Confusion matrices for the 40 ms prediction window.}
\label{tab:confusion}
\centering
\small
\begin{tabular}{lcc}
\hline
\multicolumn{3}{c}{\textbf{Complete feature model}} \\
\hline
 & Pred. Non Disruptive & Pred. Disruptive \\
\hline
Actual Non Disruptive & 103 & 25 \\
Actual Disruptive & 29 & 79 \\
\hline
\multicolumn{3}{c}{}\\[-2mm]
\multicolumn{3}{c}{\textbf{Reduced feature model}} \\
\hline
 & Pred. Non Disruptive & Pred. Disruptive \\
\hline
Actual Non Disruptive & 105 & 23 \\
Actual Disruptive & 28 & 80 \\
\hline
\end{tabular}
\end{table}

\subsection{Discussion}
The feature selection results provide important insight into the physical processes governing disruption development during the early phase of ADITYA discharges. Although thirty statistical descriptors were initially extracted from the six diagnostic signals, the DT analysis consistently identified only a limited subset of features as being most relevant for disruption discrimination. The persistence of specific diagnostics and statistical descriptors across multiple observation windows suggests that disruption precursors are embedded within distinct plasma behaviors that evolve throughout the startup phase.
A notable observation is the frequent selection of plasma current descriptors, particularly the mean, variance and wavelet energy entropy. The mean plasma current appears as an important feature in almost all observation windows, indicating that the overall current evolution during startup differs systematically between disruptive and non disruptive discharges. As the observation window increases, the variance and wavelet energy entropy of plasma current also become significant, suggesting that fluctuations and multiscale temporal variations in current profile contain additional information related to disruption onset. This behavior is physically consistent with the gradual development of current profile irregularities and MHD activity that precede plasma termination \cite{li2023mhd}.
Among the radiation diagnostics, the SXR signal emerges as one of the most consistently selected diagnostics. The mean SXR level is retained from the shortest observation windows up to 40 ms, while variance and wavelet energy entropy become increasingly important at longer windows. Since SXR emission is closely linked to core plasma conditions, impurity accumulation and energy confinement \cite{chattopadhyay2009instability, de1981soft}. The prominence of SXR based features indicates that disruption precursors are reflected in changes to core plasma behavior at a very early stage of discharge evolution.
The HXR diagnostics also contribute strongly to disruption prediction, particularly through variance, kurtosis and wavelet energy entropy. Unlike mean based descriptors, these higher-order statistical measures are sensitive to intermittent bursts and non Gaussian fluctuations associated with runaway electron generation and energetic particle dynamics \cite{purohit2020lanthanum,shevelev2021study}. Their repeated selection suggests that disruptive discharges exhibit increasingly irregular HXR activity, even within the first 20-50 ms of plasma evolution.

Features derived from the C-III emission signal are another recurring component of the reduced feature set. The mean C-III intensity remains important across nearly all observation windows, while skewness and kurtosis become significant in the later stages. Since C-III radiation is associated with impurity behavior and edge plasma conditions \cite{chowdhuri2022impurity}, these results indicate that impurity related processes contribute to disruption development from the earliest phases of discharge formation. The increasing importance of higher order moments at longer windows further suggests that the impurity dynamics become progressively more asymmetric and nonlinear as the plasma evolves.
In contrast, $H\alpha$ descriptors appear less frequently but become important through variance, skewness and wavelet energy entropy rather than through the mean value. This observation implies that edge recycling and particle source fluctuations influence disruption development primarily through transient events rather than through changes in the average emission level. The appearance of $H\alpha$ wavelet energy entropy in both the shortest and longest observation windows highlights the multiscale nature of edge plasma fluctuations during disruption evolution \cite{dey2019modeling}.
Bolometer derived features exhibit a clear transition with increasing observation time. Variance dominates the shorter windows, whereas skewness, kurtosis and wavelet energy entropy become more important beyond 35 ms. Because bolometer signals represent total radiated power losses \cite{tahiliani2009radiation,pandya2012development,meister2019bolometer}, this trend suggests that disruptions initially manifest through changes in overall radiation fluctuation levels before evolving into more complex radiative events characterized by intermittent bursts and nonlinear behavior.

The evolution of the selected statistical descriptors also reveals a systematic progression in the nature of disruption precursors. In the earliest observation windows (0-20 ms and 0-25 ms), feature selection is dominated by mean and variance, indicating that global trends and fluctuation amplitudes contain the primary discriminative information. As the observation window expands, kurtosis, skewness and wavelet energy entropy appear with increasing frequency. These descriptors are sensitive to distribution asymmetry, intermittency and multiscale fluctuations, respectively. This suggests that disruption precursors evolve from relatively simple macroscopic variations to increasingly complex nonlinearity. This transition is consistent with the progressive growth of MHD instabilities, impurity radiation events and energetic particle activity during the approach to disruption \cite{purohit2020characterization,dhyani2014electrode}.

\section{\label{sec:level1}Conclusion and Future Work}
This work presents an interpretable ML framework for early disruption prediction in the ADITYA tokamak using statistical descriptors extracted from multiple plasma diagnostic signals. Unlike conventional disruption prediction approaches that rely on raw diagnostic time series or deep learning models, the proposed methodology combines statistical feature engineering, DT based feature selection and RF classification to achieve both predictive performance and physical interpretability.
The DT analysis revealed that only a compact subset of statistical descriptors contributes significantly to disruption discrimination and that the dominant precursor features evolve with the observation window. Lower order statistical moments, particularly the mean and variance are found to be most informative during the earliest phase of plasma evolution, whereas higher order moments and wavelet energy entropy became increasingly important as additional plasma evolution was included. These observations suggest that disruption precursors evolve from relatively simple macroscopic variations to more complex nonlinear fluctuation patterns.
Using these statistically derived features, the RF classifier achieved reliable prediction performance across all investigated observation windows. The highest predictive capability was obtained for the 0-35 ms and 0-40 ms observation windows, with a maximum ROC-AUC of 0.87, demonstrating that meaningful disruption precursor information is already available within the first 20-50 ms of plasma evolution. Importantly, the reduced feature representation obtained through DT based feature selection achieved performance comparable to and in some cases marginally better than, that obtained using the complete feature set. This indicates that redundant statistical descriptors can be eliminated without sacrificing predictive accuracy, while simultaneously improving model interpretability and reducing computational complexity.
The proposed framework addresses the specific challenges associated with short pulse tokamaks, where disruption prediction must be performed using only a limited interval of plasma evolution. In contrast to recent interpretable ML studies developed for long pulse devices such as EAST, the present work demonstrates that reliable and explainable disruption prediction can be achieved during the early phase of short pulse ADITYA discharges using compact statistical representations of routinely measured diagnostic signals. The methodology therefore provides an efficient and physically interpretable alternative to raw time series learning for disruption prediction.
Future work will focus on integrating the proposed framework into an real-time disruption monitoring system, evaluating end to end inference latency and validating its robustness using data from the upgraded ADITYA-U tokamak across a broader range of plasma operating conditions. These developments will help establish the proposed methodology as a practical framework for real time disruption prediction and mitigation in present and future short pulse magnetic confinement devices.

\section*{Acknowledgment}{We acknowledge the support of the ADITYA Tokamak team, Institute for Plasma Research, India, for providing the necessary data for this study.}




\bibliographystyle{abbrv}
\bibliography{article}
\end{document}